\documentclass[preprint,12pt]{elsarticle}

\usepackage{lineno}
\usepackage{amssymb}
\usepackage{hyperref}
\usepackage{subcaption}
\usepackage{xcolor}
\journal{Nuclear Inst. and Methods in Physics Research, A}

\begin{document}
\begin{frontmatter}

\title{Performance of a spaghetti calorimeter prototype with tungsten absorber and garnet crystal fibres}

\author[CERN]{L.~An}
\author[CERN]{E.~Auffray}
\author[CERN]{F.~Betti}
\author[CERN]{F.~Dall'Omo}
\author[ICCUB]{D.~Gascon}
\author[ICL]{A.~Golutvin}
\author[IHEP]{Y.~Guz}
\author[IHEP,MISIS]{S.~Kholodenko}
\author[CERN,UNIMIB]{L.~Martinazzoli\corref{cor}}
\author[Val]{J.~Mazorra~De~Cos}
\author[ICCUB]{E.~Picatoste}
\author[CERN,UNIMIB]{M.~Pizzichemi}
\author[CERN]{P.~Roloff}
\author[CERN]{M.~Salomoni}
\author[ICCUB]{D.~Sanchez}
\author[CERN]{A.~Schopper}
\author[ITEP]{A.~Semennikov}
\author[ITEP]{P.~Shatalov}
\author[MISIS]{E.~Shmanin}
\author[MISIS]{D.~Strekalina}
\author[PKU]{Y.~Zhang}

\affiliation[CERN]{organization={European Organization for Nuclear Research (CERN)},
            city={Geneva},
            country={Switzerland}}
\affiliation[ICL]{organization={Imperial College London},
            city={London},
            country={United Kingdom}}
\affiliation[IHEP]{organization={Institute for High Energy Physics NRC Kurchatov Institute (IHEP~NRC~KI)},
            city={Protvino},
            country={Russia}}
\affiliation[ITEP]{organization={Institute of Theoretical and Experimental Physics NRC Kurchatov Institute (ITEP~NRC~KI)},
            city={Moscow},
            country={Russia}}
\affiliation[Val]{organization={Instituto de Fisica Corpuscular, Centro Mixto Universidad de Valencia~CSIC},
            city={Valencia},
            country={Spain}}
\affiliation[MISIS]{organization={National University of Science and Technology “MISIS”},
            city={Moscow},
            country={Russia}}
\affiliation[PKU]{organization={School of Physics State Key Laboratory of Nuclear Physics and Technology, Peking University},
            city={Beijing},
            country={China}}
\affiliation[UNIMIB]{organization={Universit\'a degli Studi di Milano-Bicocca},
            city={Milano},
            country={Italy}}
\affiliation[ICCUB]{organization={Universitat de Barcelona (ICCUB)},
            city={Barcelona},
            country={Spain}}
\cortext[cor]{Corresponding author: loris.martinazzoli@cern.ch}

\begin{abstract}
A spaghetti calorimeter (SPACAL) prototype with scintillating crystal fibres was assembled and tested with electron beams of energy from 1 to 5~GeV. The prototype comprised radiation-hard Cerium-doped Gd$_3$Al$_2$Ga$_3$O$_{12}$ (GAGG:Ce) and Y$_3$Al$_5$O$_{12}$ (YAG:Ce) embedded in a pure tungsten absorber.
The energy resolution was studied as a function of the incidence angle of the beam and found to be of the order of   $10\% / \sqrt{E} \oplus1\%$, in line with the LHCb Shashlik technology.
The time resolution was measured with metal channel dynode photomultipliers placed in contact with the fibres or coupled via a light guide, additionally testing an optical tape to glue the components.
Time resolution of a few tens of picosecond was achieved for all the energies reaching down to (18.5~$\pm$~0.2)~ps at 5~GeV.
\end{abstract}


\end{frontmatter}

\section{Introduction}
\label{sec:intro}
Electromagnetic calorimeters (ECAL) are designed to provide energy, position and potentially time measurements for electrons, positrons, and photons. Amongst them is the upgraded LHCb ECAL, wherefrom stems the R\&D here presented.

LHCb is an experiment dedicated to heavy flavour physics at the LHC~\cite{Collaboration_2008}. Its primary goal is to look for indirect evidence of new physics in CP violation and rare decays of beauty and charm hadrons. The LHCb ECAL is a 7.76$\times$6.30~m$^2$ wall 12.6~m downstream of the interaction point, made of Shashlik modules~\cite{caloLHCb, caloLHCbPerf} in a non-pointing layout, with an angular acceptance from 25~mrad (1.43$^{\circ}$) to 250~mrad (14.3$^{\circ}$) vertically and to 
300~mrad (17.2$^{\circ}$) horizontally. During the High-Luminosity (HL) phase of the LHC, the LHCb ECAL will face a considerable increase in particle density, especially at lower angles close to the beam-pipe~\cite{FTDR}. There, the modules will require a completely new design, implementing radiation-hard 
materials to sustain doses up to 1~MGy, a granularity as low as 15~mm, and precise timing information of a few tens of ps for high-energy electromagnetic showers 
, while keeping the energy resolution at the current level of 10\% sampling term and 1\% constant term~\cite{Arefev:2008}. The baseline solution is a spaghetti calorimeter (SPACAL).

A SPACAL is a sampling calorimeter wherein scintillating fibres are inserted into a dense absorber. The scintillating fibres convert the deposited energy into light and transport it to the photodetectors, avoiding wavelength-shifters which entail a reduction in light collection efficiency.
At the same time, the electromagnetic shower dimension can be tuned by selecting absorber materials with adequate radiation length and Molière radius.

This paper studies a SPACAL with radiation-hard crystal fibres and tungsten absorber.
Proof of concepts of this technology obtained promising results in the past years~\cite{Lucchini_2013, martinazzoli2020}, but for the first time a novel prototype was assembled with pure tungsten absorber, garnet crystal fibres, and photomultipliers developed for picosecond timing. It was tested  in the TB24 area of the DESY II testbeam facility 
with 1 to 5~GeV electron beams~\cite{Diener:2018qap}, measuring the energy resolution at different incidence angles of the beam and studying the time resolution with particular emphasis on the role of optical coupling.

Materials and solutions employed here can be relevant for other future High Energy Physics (HEP) experiments. For example, the achieved energy and time resolutions in combination with the radiation hardness make it an interesting option for experiments at a future hadron collider \cite{FCC:2018vvp} and fixed-target experiments at the intensity frontier \cite{Alemany:2019vsk}.

\section{SPACAL Prototype}
\label{sec:prototype}
\begin{figure}[t]
\centering
    \subfloat[\label{fig:SPACALProto_1}\centering Front view]{{\includegraphics[width=0.43\textwidth]{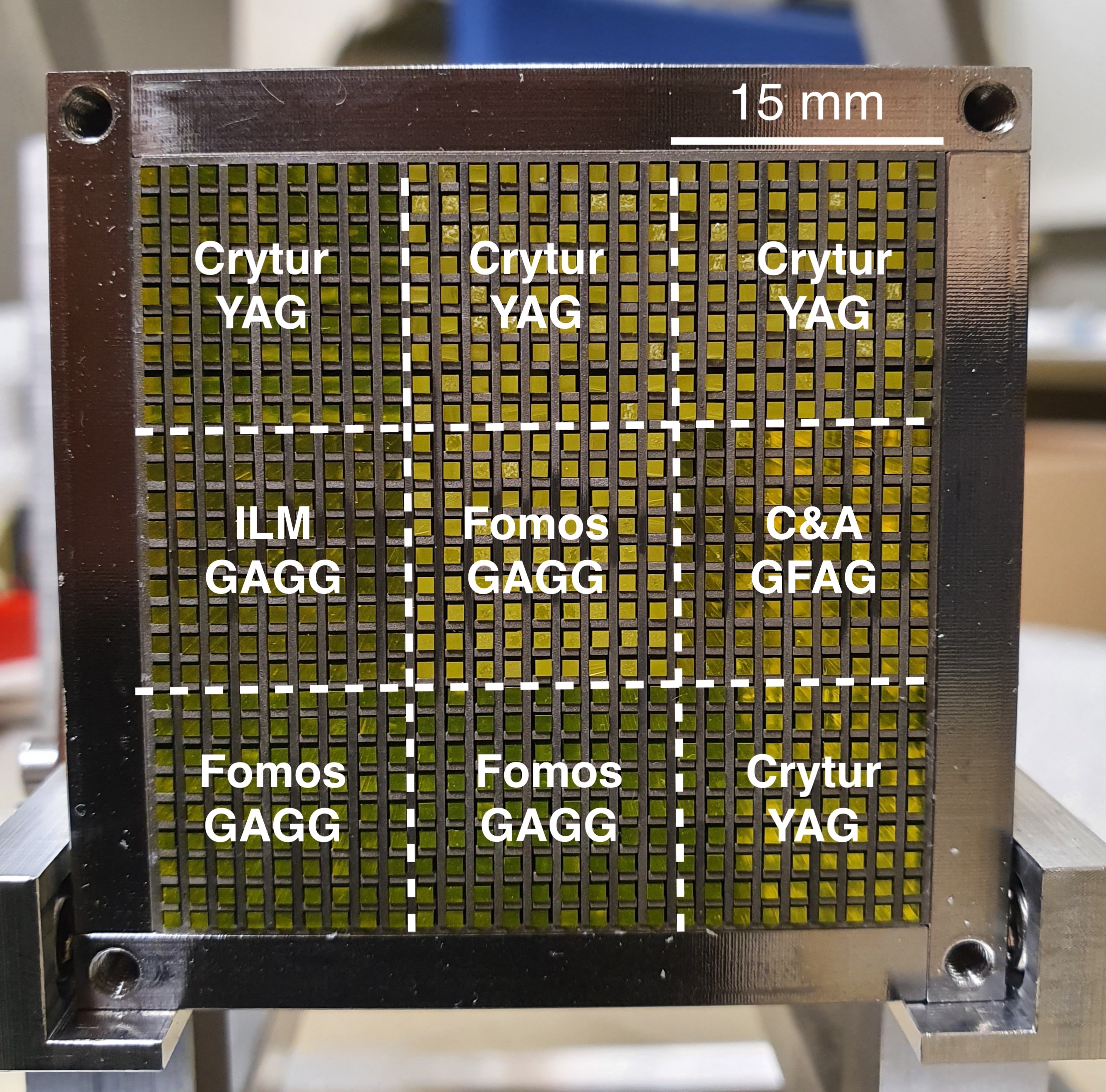}}}%
    \,
    \subfloat[\label{fig:SPACALProto_2}\centering Three-quarter view]{{\includegraphics[width=0.43\textwidth]{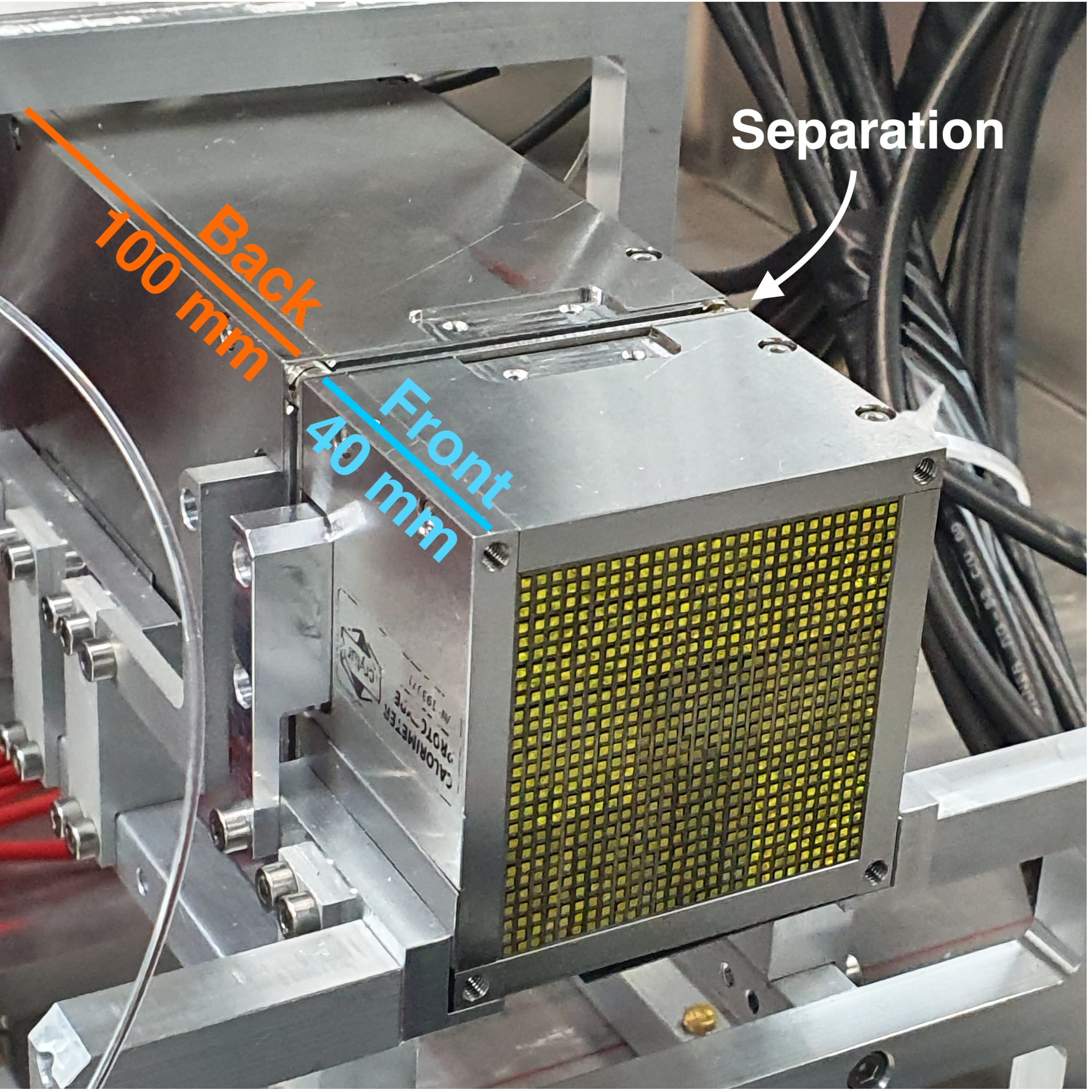}}}%
    \caption{Pictures of the Tungsten-Crystal SPACAL prototype without readout. On the left, the 9 cells with different garnets and producers are highlighted. On the right, the front and the back sections are visible, with the small gap between the two and the mechanics holding the prototype.}%
    \label{fig:SPACALProto}%
\end{figure}

The prototype assembled is a SPACAL with a pure Tungsten absorber and garnet crystal fibres (Fig.~\ref{fig:SPACALProto}). It is longitudinally segmented into a front and a back section close to the average shower maximum. The two were made independent to allow inserting a large area picosecond photodetector (LAPPD) as an optional timing layer between the two, discussed in~\cite{Perazzini:2022}. 

Each section is divided into 9 cells of 15$\times$15~mm$^2$ surface, read out by photomultipier tubes (PMT) discussed in the measurement sections.

\subsection{The Scintillating Fibres}
The scintillating fibres are Cerium-doped garnet crystals, namely \newline Gd$_3$Al$_2$Ga$_3$O$_{12}$ (GAGG) and Y$_3$Al$_5$O$_{12}$ (YAG), which are known for their radiation hardness up to several hundreds of kGy~\cite{GAGGRadHard,GarnetsRadHard}. 
The GAGG crystals were selected to achieve the best timing performance following a characterisation campaign involving several producers~\cite{martinazzoli2021}.  
Each section of the prototype was equipped with 1 cell of GAGG from
Institut Lumière Matière (ILM, France), 3 of GAGG from Fomos (Russia), 1 of Gadolinium Fine Aluminum Gallate (GFAG), a commercial name for the fast-timing GAGG by C\&A (Japan), and the remaining 4 with YAG fibres from Crytur spol. s.r.o. (Czech Republic). The fibres were arranged in the cells as shown in Fig.~\ref{fig:SPACALProto}, with the same crystal type in the front and in the back sections.

\subsection{The Absorber} 
The absorber, produced by Crytur, is made of pure Tungsten plates 0.5~mm thick and 19~g/cm$^{3}$ dense. 
Rows of 0.5~mm were carved with a pitch of 1.7~mm in the planes for half of their length, resembling a comb. Half of the planes could be inserted into the other half rotated by $90^{\circ}$, thus forming a grid of rows with 0.5~mm thick tungsten walls and squared holes of 1.2~mm side, wherein the fibres could be inserted. Therefore, the pitch between adjacent fibres is 1.7~mm.
The front and the back sections are 40~mm and 100~mm long, respectively. With a radiation length X$_0$ of 6~mm, the first section accounts for about 7~X$_0$, corresponding to the position of the average shower maximum for 20~GeV electrons.  The Molière radius is 14.3~mm and 15.2~mm for the GAGG and YAG cells, respectively, computed as the volume-weighted average of the materials' radii. 
The shower containment predicted by Monte Carlo simulations (see Sec.~\ref{sec:MonteCarloSims}) for electrons hitting the central cell is above 90\%.
The front section was read out from the front, the back one from the back. On the side opposite to the readout, a thin 3M Enhanced Specular Reflector (ESR) film layer was pressed against the fibres by a stainless steel plate to reflect the light towards the readout. These plates were 1.50~mm and 1.05~mm thick for the front and the back sections, respectively.

\section{Testbeam Setup}
\label{sec:tb_setup}

\begin{figure}[t]
\centering
\includegraphics[width=1\textwidth]{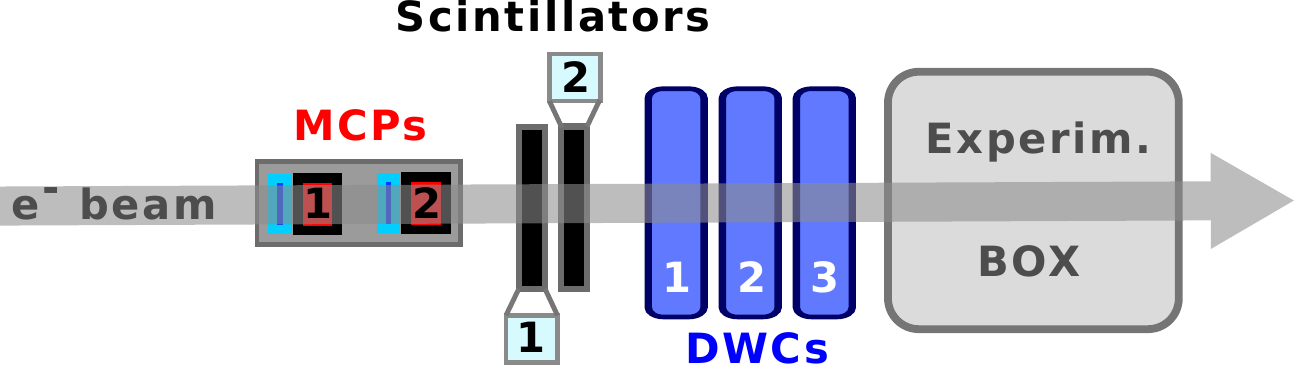}
\caption{Test beam setup. The electron beam moves from left to right, two MCPs provide the time reference, two scintillating pads the trigger signal, and three DWCs the tracking information. The experimental box contains the prototype and the rotating steppers.}
\label{fig::tbSETUP}
\end{figure}

The testbeam setup is sketched in Fig.~\ref{fig::tbSETUP}.
The prototype was mounted on two steppers allowing it to rotate horizontally (yaw) and vertically (pitch).
The rotation of the prototype is defined as yaw~+~pitch hereafter, e.g. $3^{\circ}$+$3^{\circ}$.
The assembly was placed inside a light-tight box and aligned with the beam using a laser system supplied by the facility. The box was then installed on a table moving 
in the plane
orthogonal to the beam. 
The uncertainty on the incidence angle was estimated to be $\pm 0.15^{\circ}$ both horizontally and vertically.

Cables of 0.8~m in length were employed inside the box to connect the prototype to a patch panel on the box surface, and 3~m long ones from the patch panel to the electronics rack. Additionally, the time resolution was studied replacing the 3~m long cables with 12~m long ones.

For the energy measurements, the signals were integrated over a 400~ns gate using 3 LeCroy 1182 ADC modules.
For the timing measurements, the waveforms were digitised using the DRS4-based~\cite{Ritt:DRS4} V1742 CAEN digitiser with 5 Gs/s and 500 MHz bandwidth, with a custom calibration based on~\cite{ShaverRitt:DRS4}. 
A 10~dB electronic attenuator was employed in order to match the PMTs' pulse heights to the dynamic range of the digitiser, except for the measurements with 10~cm light guides.

The hardware trigger was provided by 2 plastic scintillating pads in coincidence. Tracking information was given by 3 delay wire chambers (DWCs) employing a mixture of Ar/CO$_2$ gas and read out by a CAEN TDC V1290N. 
The time reference was given by 2 microchannel plate detectors (MCP) along the beam line. 
The timestamp of each MCP was computed by constant fraction discrimination (CFD) at 30\% on the digitised waveform and the average of the 2 timestamps was used as time reference. The time resolution of the reference was measured run by run with the two MCPs in coincidence. The typical resolution was 14~ps (standard deviation), stable throughout the data acquisition campaign.

\section{Data Analysis and Results}
\label{sec:results}



\subsection{Energy resolution}
The energy resolution was measured by coupling all the 18 cells to Hamamatsu R12421 PMTs via 30~mm long PMMA light guides in dry contact, i.e. without optical grease or glue.

\subsubsection{Calibration}
\label{sec:EResCalib}
The prototype was tilted by 1$^{\circ}$+1$^{\circ}$ with respect to a 3~GeV beam. First, the front PMTs' bias voltages were tuned to achieve the same peak position of the charge histograms of each cell with the electron beam hitting the centre of that cell. Afterwards, the front section was removed and the same procedure was repeated for the back section.
Calibrating both front and back sections simultaneously with only electromagnetic showers would increase the response non-linearity \cite{GANEL1998621}. Therefore, the back section was calibrated stand-alone, i.e. removing the front one, being sufficiently deep to achieve a reasonable containment of 3~GeV electron showers.
Selecting only events hitting in a 20$\times$20~mm$^2$ square in the centre of the back section, a set of 9 calibration coefficients $c_j$ for the back cells was found by minimising the residuals~\cite{martinazzoli2020}:
\begin{equation}
    \sum_{i_{ev} = 1}^{N_{ev}} \left[E_{beam} - \sum_{j=1}^{9} c_{b,j} S_{b,j} \right]^2 = \mbox{min},
\end{equation}
where $S_{b,j}$ and $c_{b,j}$ are the integrated charge and the calibration coefficient of the $j$-th cell in the back section, $E_{beam}$ equals 3~GeV and $i_{ev}$ the $i$-th event of the dataset. Afterwards, the front section was reinstalled, and the procedure was repeated for the whole module keeping the calibration coefficients of the back constant:
\begin{equation}
    \sum_{i_{ev} = 1}^{N_{ev}} \left[ E_{beam} - \sum_{j=1}^{9} \left(c_{f,j} S_{f,j} + c_{b,j} S_{b,j}\right) \right]^2 = \mbox{min},
\end{equation}
where $S_{f,j}$ and $c_{f,j}$ are the integrated charge and the calibration coefficient of the $j$-th cell in the front section.


\subsubsection{Energy Measurements}
\label{sec:EResMeas}
\begin{figure}[t]
\centering
\includegraphics[width=.95\textwidth]{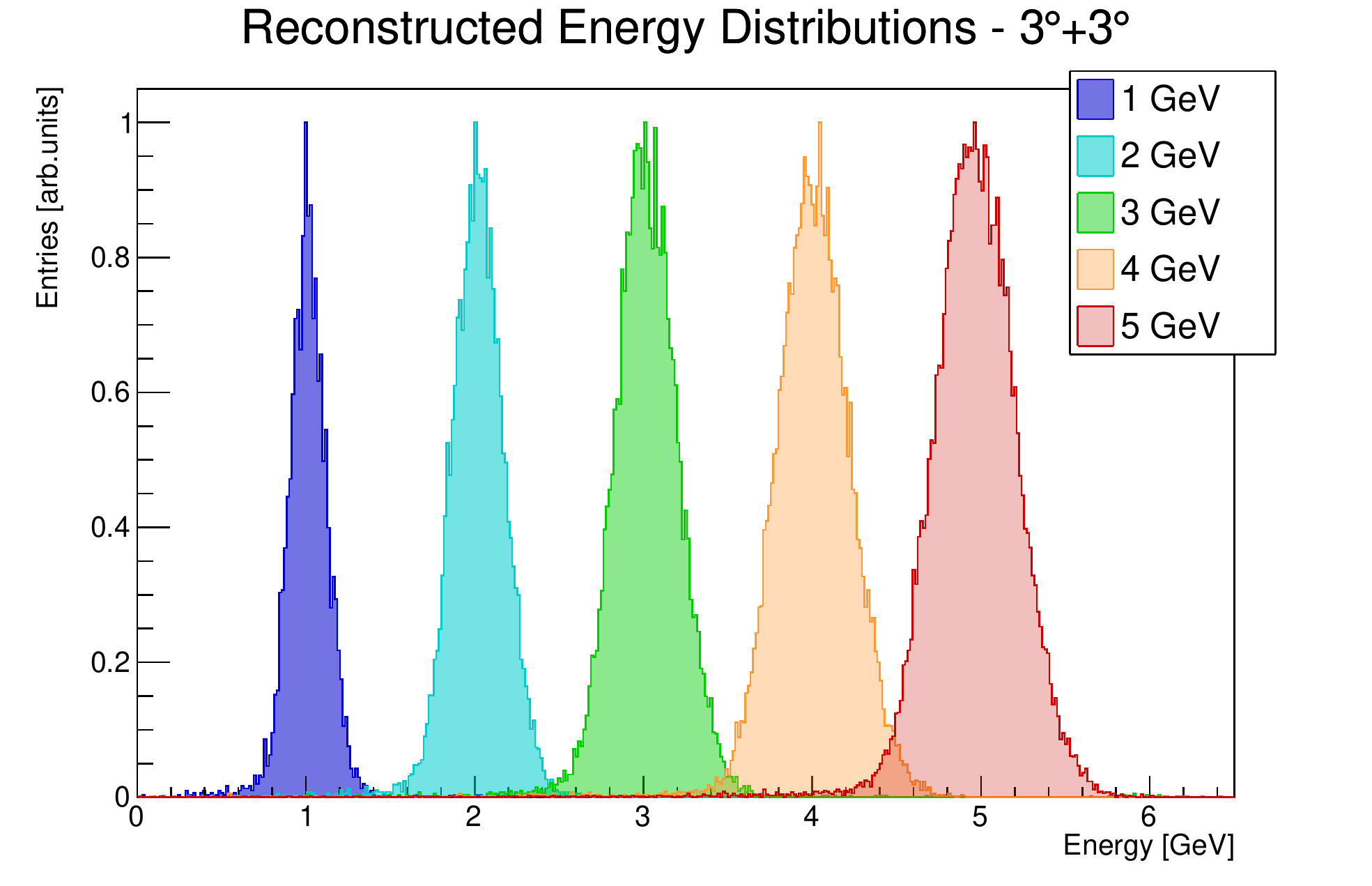}
\caption{Distributions of the reconstructed energy at 3$^{\circ}$+3$^{\circ}$ incidence angle rescaled to the same height.}
\label{fig::energyDistro}
\end{figure}

\begin{figure}[t]
\centering
\includegraphics[width=.97\textwidth]{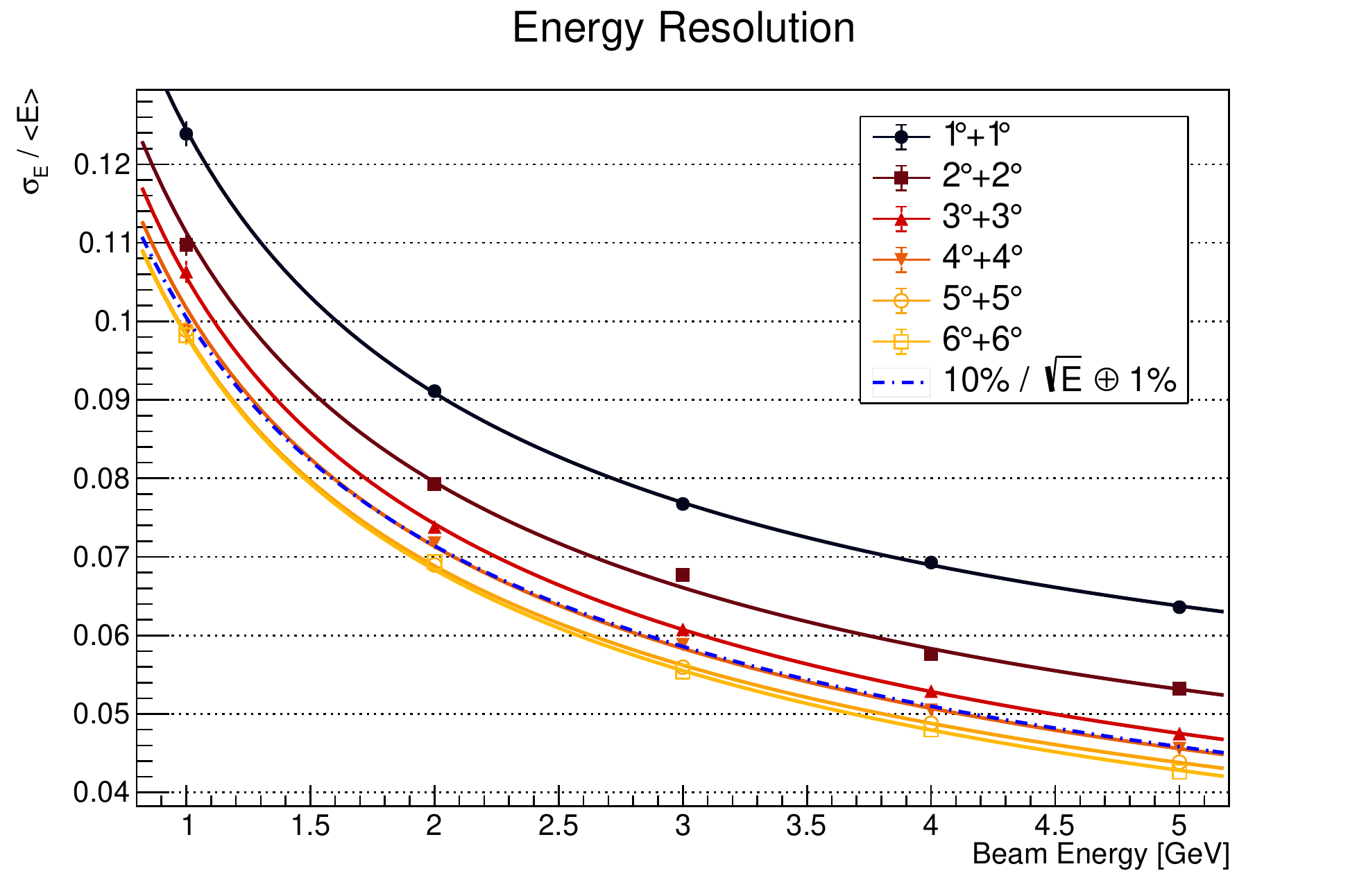}
\caption{Energy resolution of the prototype measured at different incidence angles of the electron beam. 
The lines are fits of Eq.~\ref{eq:energyRes} to the datapoints. The common noise term is (0.024~$\pm$~0.004)~GeV. The resolution improves with increasing incidence angle. This effect fades above 3$^{\circ}$+3$^{\circ}$. 
The blue dotted line is the resolution reached with 10\% sampling and 1\% constant terms.}
\label{fig::energyRes}
\end{figure}

The energy measurements were carried out with electrons of 1 to 5~GeV hitting the centre of the prototype in a 5$\times$5~mm$^2$ square. 
Between 5000 and 25000 events were used for each energy and angle.
The reconstructed energy was computed summing the integrated charge of the 18 cells weighted by the calibration coefficients 
(see Sec. \ref{sec:EResCalib}).
The distributions of reconstructed energy at 3$^{\circ}$+3$^{\circ}$ incidence angle are visible in Fig.~\ref{fig::energyDistro}. These were fitted with a Gaussian function whose standard deviation divided by its mean is the energy resolution. The Gaussian fit describes well the energy distributions for incidence angles greater than 1°+1°.
The non-linearity over the range tested was within $\pm$1\% and it will be subject of future studies with higher-energy beams.

The measured resolutions (see Fig.~\ref{fig::energyRes}) against the beam energy were described by a sampling $s$, a constant $c$, and a noise $n$ term:
\begin{equation}
\label{eq:energyRes}
    \frac{\sigma}{E} = \frac{s}{\sqrt{E}} \oplus \frac{n}{E} \oplus c,
\end{equation}
where $\oplus$ is a sum in quadrature and $E$ is the beam energy in GeV. The data were fitted with Eq.~\ref{eq:energyRes} employing one sampling and one constant term per incidence angle and one common noise term for the six angles, which was found to be (0.024~$\pm$~0.004)~GeV.

The energy resolution improves increasing the incidence angle, with the effect fading above 3$^{\circ}$+3$^{\circ}$. At 3$^{\circ}$+3$^{\circ}$ and 6$^{\circ}$+6$^{\circ}$ the sampling terms are (10.2~$\pm$~0.1)\% and (9.5~$\pm$~0.1)\%
, respectively.
The shower is transversally narrower than the fibres pitch at the beginning of its development, giving rise to large differences in energy deposit and longitudinal fluctuations in its start position depending on whether the primary electron first hits the absorber or a fibre, as already observed in \cite{AKCHURIN200529}.
Tilting the prototype offers a higher sampling rate along the direction of the primary electron, thus reducing the above fluctuations and improving energy resolution. 

The constant term and its statistical error at 3$^{\circ}$+3$^{\circ}$ are found to be (1.2~$\pm$~0.3)\%. Several potential sources of systematic uncertainty were investigated. 
First, the impact on the constant and sampling terms of a misalignment up to $0.15^{\circ}$ (see Sec.~\ref{sec:tb_setup}) is smaller than the statistical uncertainty.
Secondly, the momentum spread of the beam is unknown, but expected to be much smaller than the 158~MeV/c measured for the TB21 area \cite{Diener:2018qap} due to the configuration of the TB24 beamline; for illustration, an uncertainty of 50~MeV on the beam energies added to the fit doubles the statistical uncertainties on the constant terms.
As a third test, the constant terms at 3$^{\circ}$+3$^{\circ}$ and larger incidences were assumed to be up to 2\% and a fit to extract the sampling and noise terms was performed; compared to the nominal procedure, the sampling terms decreased maximally by 1\%.
Finally, a variation on the calibration method was tested applying a weight to the back section's calibration factors to minimise the non-linearity over the 1-5~GeV range instead of the resolution; the constant term at 3$^{\circ}$+3$^{\circ}$ increases to (1.9~$\pm$~0.2)\% and similar values are found at larger angles, whereas the sampling terms are unaffected within the statistical uncertainty. 
In conclusion, whilst the sampling terms can be reliably determined in the energy range available, a precise measurement of the constant terms and their systematic uncertainties requires data at higher energies.


\subsubsection{Monte Carlo Simulations}
\label{sec:MonteCarloSims}
\begin{figure}[t]
\centering
\includegraphics[width=.9\textwidth]{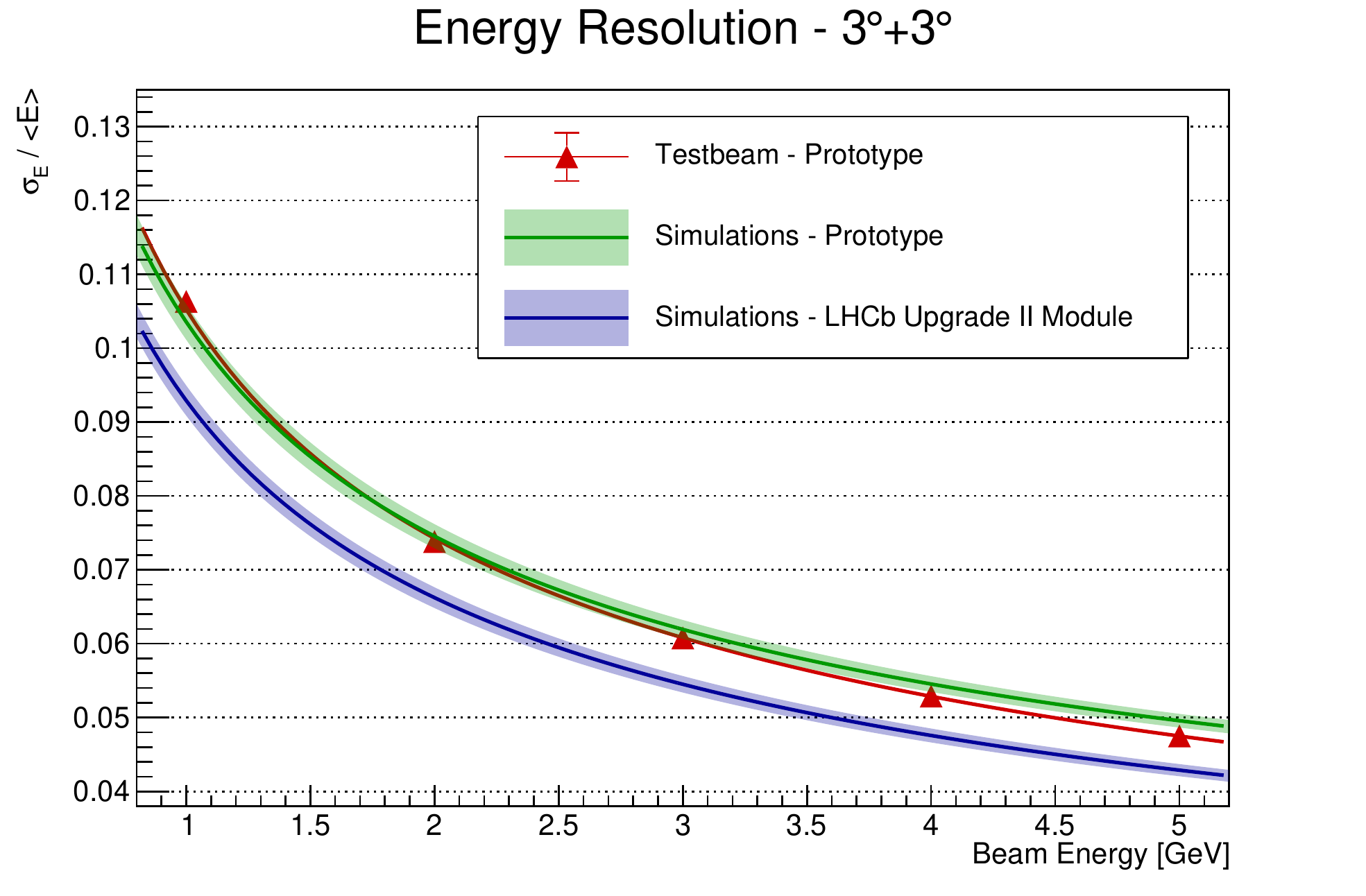}
\caption{Energy resolution to electrons at 3$^{\circ}$+3$^{\circ}$ incidence angle. Comparison between testbeam measurements of the prototype, simulations of the prototype, and simulations of the module designed for the LHCb Upgrade II \cite{FTDR}. Simulations are drawn with a 2 standard deviations error band.}
\label{fig::energyResSims}
\end{figure}

To deepen the understanding of the contributions to the resolution, the prototype was implemented in the GEANT4 Monte Carlo simulation framework~\cite{AGOSTINELLI2003250}. 
Fig.~\ref{fig::energyResSims} shows the agreement between the simulated energy resolution at 3$^{\circ}$+3$^{\circ}$ incidence, as a green curve with a 2 standard deviations error band, and the measurements.

The framework was then employed to simulate a calorimeter module of 120$\times$120~mm$^2$ with an optimised design as described in the LHCb Framework Technical Design Report \cite{FTDR} 
-- namely with 45/105~mm long sections, without stainless steel plates between the two, and with 1.67~mm pitch --
achieving sampling and constant terms of (9.2~$\pm$~0.1)\% and (1.18~$\pm$~0.03)\%, respectively (blue curve in Fig.~\ref{fig::energyResSims}).

\subsection{Time resolution}

\begin{figure}[t]
\centering
\includegraphics[width=.9\textwidth]{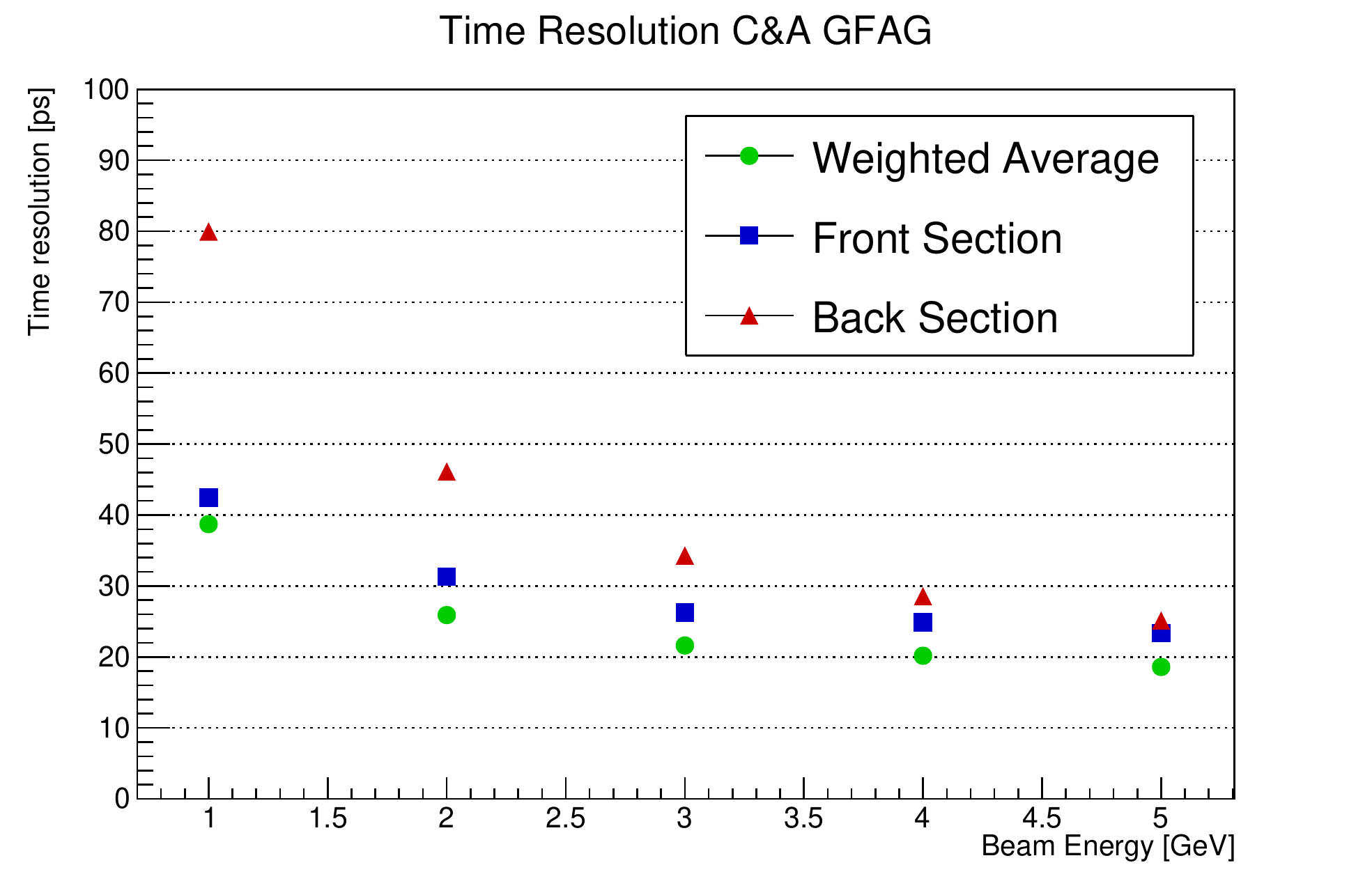}
\caption{Time resolution of the GFAG cells with Hamamatsu MCD R7600U-20 PMTs in direct dry contact. The three set of resolutions are obtained using the front cell, the back one, and the inverse-variance weighted average of the two timestamps.}
\label{fig:wGFAGMethods}
\end{figure}
The time measurements were carried out instrumenting 
1 front and 1 back cell with Hamamatsu R7600U-20 metal channel dynode (MCD) PMTs. 
The MCD technology ensures better time resolution due to the short path and small spatial spread of the electron cloud. The PMTs were placed in direct dry contact with the prototype cell and the excess photocathode's surface was masked to avoid collecting light from the neighbouring ones. 
Approximately 2000 events were used per data point.
For each event, the timestamp of a cell was computed applying CFD to the digitised waveform; the optimal threshold was found with a scan to be 15\%. 
All the measurements were performed with a beam incidence angle of 3$^{\circ}$+3$^{\circ}$ and selecting tracks hitting the centre of the front cell in a 4$\times$4~mm$^2$ square. The difference between a cell timestamp and the time reference was computed and the distributions of this difference for each beam energy were then fitted with a Gaussian function whose standard deviation, after subtracting quadratically the reference's contribution, is the time resolution. The sources of systematic errors discussed in Sec.~\ref{sec:EResMeas} were found negligible with respect to the statistical uncertainties for the timing measurements.

Fig.~\ref{fig:wGFAGMethods} shows the time resolution of the GFAG cells. The front one provides better time resolution than the back one, with the spread between the two diminishing with increasing energy.
This is expected for electrons of a few GeV which produce a relatively short shower and deposit on average the larger fraction of their energy in the front cell.
The time resolution reaches down to (18.5~$\pm$~0.2)~ps at 5~GeV using as timestamp the weighted average of the timestamps of the 2 cells and as weights the inverse of their variance at that energy, i.e. $\frac{1}{\sigma_t^2(E)}$. 

In agreement with the laboratory tests~\cite{martinazzoli2021}, the fibres from ILM and C\&A show a similar resolution, whereas those from Fomos are a few ps worse (see later Fig. \ref{fig:wGAGGCoupl} and Sec. \ref{Sec:OpticalCoupling}). No degradation of time resolution within the experimental uncertainty is visible when using 12.8~m instead of 3.8~m of total cable length to connect the PMTs to the digitiser.

\subsubsection{Time resolution and Optical Coupling}
\label{Sec:OpticalCoupling}
\begin{figure}[t]
\centering
\includegraphics[width=.9\textwidth]{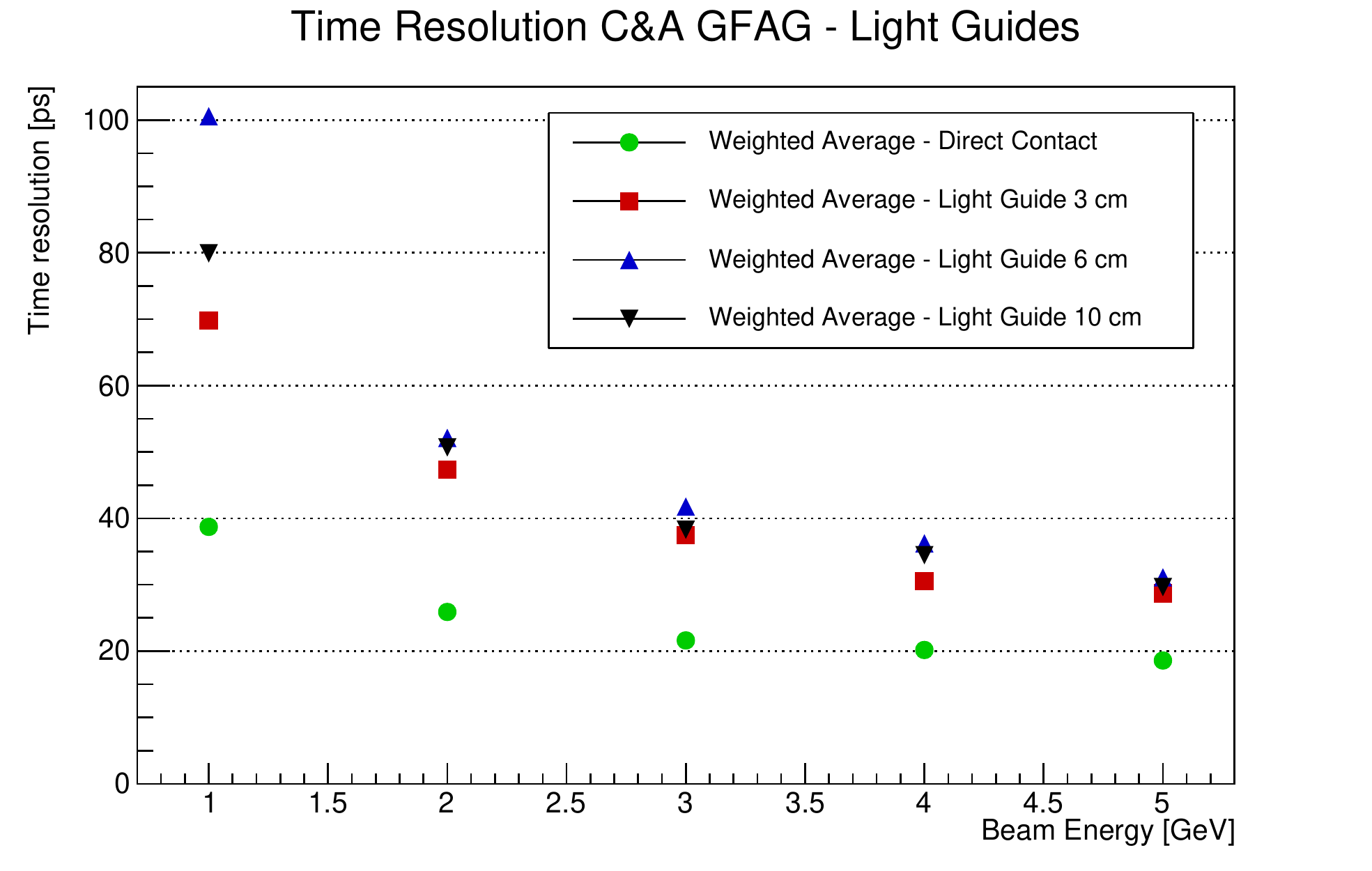}
\caption{Time resolution of C\&A GFAG combining the front and the back cells in direct dry contact with the MCD PMTs or using PMMA light guides. Employing a light guide worsens time resolution, but the deterioration depends little on the length.
}
\label{fig:wGFAGLG}
\end{figure}

\begin{figure}[t]
\centering
\includegraphics[width=.9\textwidth]{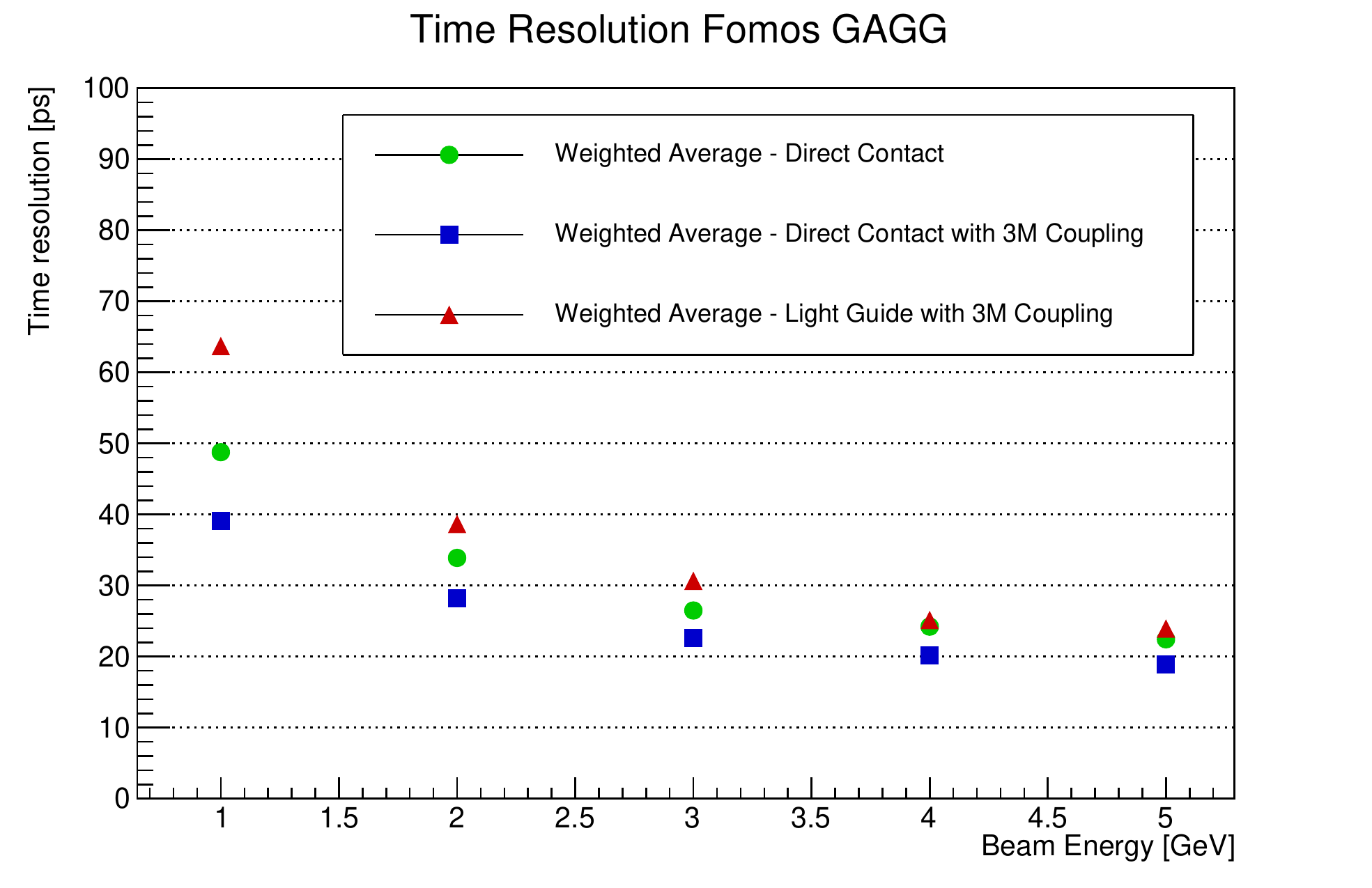}
\captionof{figure}{Time resolution of Fomos GAGG combining the front and the back cells coupled to the PMTs using a 3M optical adhesive with an index of refraction of 1.47.}
\label{fig:wGAGGCoupl}
\end{figure}

Achieving fast timing with light-based detectors requires maximising the amount of optical photons detected in the first nanoseconds of the scintillation process, i.e. the photon time-density~\cite{Gundacker_2020, VINOGRADOV2018149}.

The time resolution of the GFAG cells was compared placing the PMT in direct dry contact with the fibres or via a PMMA light guide of length 30, 60, or 100~mm (Fig.~\ref{fig:wGFAGLG}). With the light guides, the resolution is degraded by a factor 1.5 to 2 in the energy range considered, being close to 30~ps at 5~GeV. The light guides length has little influence on the resolution or on the pulse amplitude, 
therefore light absorption inside the light guides cannot account for the loss of performance. 
The discrepancy at 1~GeV between the dataset with 6 and 10~cm light guides could be explained by the lack of electronic attenuation.

Another source of light loss is given by the Fresnel reflections taking place at each optical interface between media with mismatching indices of refraction. The entrance and exit face of a light guide constitute such an interface with the surrounding air.
Gluing the light guide to the photocathode and the crystal fibres with some high-refractive-index optical coupling removes air from the optical chain, thus reducing Fresnel reflections. Moreover, the high index of refraction of GAGG crystals (almost 1.90 at the peak emission wavelength \cite{Fomos:IndexOfRefractionGAGG}) produces a light extraction cone of approximately 32$^{\circ}$ when surrounded by air; raising the index of refraction at the crystal end face replacing air with an optical grease widens the extraction cone and, thus, increases the light output.

The time resolution of a Fomos GAGG cell was measured in direct contact and with 30~mm long light guides using 3M optical adhesives (index of refraction 1.47) between fibres, light guides, and the PMTs (Fig.~\ref{fig:wGAGGCoupl}). In this way, the timing deterioration moving from direct dry contact to using a light guide is almost entirely mitigated. In addition to that, gluing the PMT photocathode directly to the crystal fibres improves further the time resolution, reaching better than 20~ps at 5~GeV.

\section{Conclusions}
\label{sec:conclusions}

In this work, the time and energy resolutions of a spaghetti calorimeter with 
radiation-hard
garnet crystal fibres and pure tungsten absorber were measured at the DESY II facility with electron beams of 1 to 5~GeV. 
In agreement with simulations, the prototype demonstrated an energy resolution comparable to the LHCb Shashlik modules, with a sampling and a constant term of about 10\% and 1\%, respectively. While the sampling term could be reliably estimated with the data available, the constant term will be subject of study in future testbeams at higher energies.
An excellent time resolution of a few tens of picoseconds was obtained over the energy range, reaching (18.5~$\pm$~0.2)~ps at 5~GeV. Additionally, the time resolution was studied varying the optical coupling with the readout, placing the PMTs in direct contact with the crystal fibres or coupling them via light guides, and employing optical adhesives with high index of refraction.

\section*{Acknowledgments}
\label{sec:acknowledgments}
The research leading to these results was conducted in the framework of the LHCb collaboration and the Crystal Clear Collaboration.
We acknowledge support by the CERN Strategic Programme on Technologies for Future Experiments,  https://ep-rnd.web.cern.ch/ , by the MCIN/AEI, GenCat and GVA (Spain), and by the NSFC (China) under grant Nos. 12175005, 12061141007. The measurements were performed at the Test Beam Facility at DESY Hamburg (Germany), a member of the Helmholtz Association (HGF). The authors would like to thank T.~Schneider, H.~Gerwig, N.~Siegrist, and D.~Deyrail (CERN) for their help in designing and assembling the prototype and the set-up, A. Barnyakov, Budker Institute of Nuclear Physics (BINP), Novosibirsk, for kindly providing the MCPs, and the ITEP ATLAS group for the DWCs.

\bibliographystyle{elsarticle-num} 
\bibliography{spacal}

\begin{thebibliography}{10}
\expandafter\ifx\csname url\endcsname\relax
  \def\url#1{\texttt{#1}}\fi
\expandafter\ifx\csname urlprefix\endcsname\relax\def\urlprefix{URL }\fi
\expandafter\ifx\csname href\endcsname\relax
  \def\href#1#2{#2} \def\path#1{#1}\fi

\bibitem{Collaboration_2008}
{The~LHCb~Collaboration}, {The {LHCb} Detector at the {LHC}}, Journal of
  Instrumentation 3~(08) (2008) S08005--S08005.
\newblock \href {https://doi.org/10.1088/1748-0221/3/08/s08005}
  {\path{doi:10.1088/1748-0221/3/08/s08005}}.

\bibitem{caloLHCb}
S.~Amato, et~al., \href{http://cds.cern.ch/record/494264}{{LHCb calorimeters:
  Technical Design Report}}, Technical design report. LHCb, CERN, Geneva, 2000.
\newline\urlprefix\url{http://cds.cern.ch/record/494264}

\bibitem{caloLHCbPerf}
A.~Beteta, et~al., \href{https://cds.cern.ch/record/2729028}{{Calibration and
  performance of the LHCb calorimeters in Run 1 and 2 at the LHC}}, Tech. rep.
  (Aug 2020).
\newblock \href {http://arxiv.org/abs/2008.11556} {\path{arXiv:2008.11556}}.
\newline\urlprefix\url{https://cds.cern.ch/record/2729028}

\bibitem{FTDR}
J.~Baptista~Leite, et~al., {Framework TDR for the LHCb Upgrade II}:
  {Opportunities in flavour physics, and beyond, in the HL-LHC era} (2022).

\bibitem{Arefev:2008}
A.~Arefev, et~al., {Beam test results of the LHCb electromagnetic calorimeter},
  CERN-LHCB-2007-149 (4 2008).

\bibitem{Lucchini_2013}
M.~Lucchini, et~al., {Test beam results with {LuAG} fibers for next-generation
  calorimeters}, Journal of Instrumentation 8~(10) (2013) P10017--P10017.
\newblock \href {https://doi.org/10.1088/1748-0221/8/10/p10017}
  {\path{doi:10.1088/1748-0221/8/10/p10017}}.

\bibitem{martinazzoli2020}
L.~Martinazzoli, {Crystal Fibers for the LHCb Calorimeter Upgrade}, IEEE
  Transactions on Nuclear Science 67~(6) (2020) 1003--1008.
\newblock \href {https://doi.org/10.1109/TNS.2020.2975570}
  {\path{doi:10.1109/TNS.2020.2975570}}.

\bibitem{Diener:2018qap}
R.~Diener, et~al., {The DESY II Test Beam Facility}, Nucl. Instrum. Meth. A 922
  (2019) 265--286.
\newblock \href {http://arxiv.org/abs/1807.09328} {\path{arXiv:1807.09328}},
  \href {https://doi.org/10.1016/j.nima.2018.11.133}
  {\path{doi:10.1016/j.nima.2018.11.133}}.

\bibitem{FCC:2018vvp}
A.~Abada, et~al., {FCC-hh: The Hadron Collider}: {Future Circular Collider
  Conceptual Design Report Volume 3}, Eur. Phys. J. ST 228~(4) (2019)
  755--1107.
\newblock \href {https://doi.org/10.1140/epjst/e2019-900087-0}
  {\path{doi:10.1140/epjst/e2019-900087-0}}.

\bibitem{Alemany:2019vsk}
R.~Alemany, et~al., {Summary Report of Physics Beyond Colliders at CERN} (2
  2019).
\newblock \href {http://arxiv.org/abs/1902.00260} {\path{arXiv:1902.00260}}.

\bibitem{Perazzini:2022}
S.~Perazzini, F.~Ferrari, V.~M. Vagnoni, on~behalf of~the LHCb ECAL Upgrade-2
  R\&D~Group, {Development of an MCP-Based Timing Layer for the LHCb ECAL
  Upgrade-2}, Instruments 6~(1) (2022).
\newblock \href {https://doi.org/10.3390/instruments6010007}
  {\path{doi:10.3390/instruments6010007}}.

\bibitem{GAGGRadHard}
V.~Alenkov, et~al., {Irradiation studies of a multi-doped
  Gd$_3$Al$_2$Ga$_3$O$_{12}$ scintillator}, Nucl. Instrum. Meth. A 916 (2019)
  226--229.
\newblock \href {https://doi.org/10.1016/j.nima.2018.11.101}
  {\path{doi:10.1016/j.nima.2018.11.101}}.

\bibitem{GarnetsRadHard}
M.~T. Lucchini, K.~Pauwels, K.~Blazek, S.~Ochesanu, E.~Auffray, {Radiation
  Tolerance of LuAG:Ce and YAG:Ce Crystals Under High Levels of Gamma- and
  Proton-Irradiation}, IEEE Transactions on Nuclear Science 63~(2) (2016)
  586--590.
\newblock \href {https://doi.org/10.1109/TNS.2015.2493347}
  {\path{doi:10.1109/TNS.2015.2493347}}.

\bibitem{martinazzoli2021}
L.~Martinazzoli, et~al., Scintillation properties and timing performance of
  state-of-the-art
  \textnormal{Gd}$_3$\textnormal{Al}$_2$\textnormal{Ga}$_3$\textnormal{O}$_{12}$
  single crystals, Nucl. Inst. Meth. A 1000~(165231) (2021).
\newblock \href {https://doi.org/10.1016/j.nima.2021.165231}
  {\path{doi:10.1016/j.nima.2021.165231}}.

\bibitem{Ritt:DRS4}
S.~Ritt, {Design and performance of the 6 GHz waveform digitizing chip DRS4},
  in: 2008 IEEE Nuclear Science Symposium Conference Record, 2008, pp.
  1512--1515.
\newblock \href {https://doi.org/10.1109/NSSMIC.2008.4774700}
  {\path{doi:10.1109/NSSMIC.2008.4774700}}.

\bibitem{ShaverRitt:DRS4}
D.~Stricker-Shaver, S.~Ritt, B.~J. Pichler, {Novel Calibration Method for
  Switched Capacitor Arrays Enables Time Measurements With Sub-Picosecond
  Resolution}, IEEE Transactions on Nuclear Science 61~(6) (2014) 3607--3617.
\newblock \href {https://doi.org/10.1109/TNS.2014.2366071}
  {\path{doi:10.1109/TNS.2014.2366071}}.

\bibitem{GANEL1998621}
O.~Ganel, R.~Wigmans, {On the calibration of longitudinally segmented
  calorimeter systems}, Nucl. Inst. Meth. A 409~(1) (1998) 621--628.
\newblock \href {https://doi.org/https://doi.org/10.1016/S0168-9002(97)01337-5}
  {\path{doi:https://doi.org/10.1016/S0168-9002(97)01337-5}}.

\bibitem{AKCHURIN200529}
N.~Akchurin, et~al., {Electron detection with a dual-readout calorimeter},
  Nucl. Inst. Meth. A 536~(1) (2005) 29--51.
\newblock \href {https://doi.org/https://doi.org/10.1016/j.nima.2004.06.178}
  {\path{doi:https://doi.org/10.1016/j.nima.2004.06.178}}.

\bibitem{AGOSTINELLI2003250}
S.~Agostinelli, et~al., {Geant4—a simulation toolkit}, Nucl. Inst. Meth. A
  506~(3) (2003) 250--303.
\newblock \href {https://doi.org/https://doi.org/10.1016/S0168-9002(03)01368-8}
  {\path{doi:https://doi.org/10.1016/S0168-9002(03)01368-8}}.

\bibitem{Gundacker_2020}
S.~Gundacker, et~al., {Experimental time resolution limits of modern {SiPMs}
  and {TOF}-{PET} detectors exploring different scintillators and Cherenkov
  emission}, Physics in Medicine {\&} Biology 65~(2) (2020) 025001.
\newblock \href {https://doi.org/10.1088/1361-6560/ab63b4}
  {\path{doi:10.1088/1361-6560/ab63b4}}.

\bibitem{VINOGRADOV2018149}
S.~Vinogradov, {Approximations of coincidence time resolution models of
  scintillator detectors with leading edge discriminator}, Nucl. Inst. Meth. A
  912 (2018) 149--153.
\newblock \href {https://doi.org/https://doi.org/10.1016/j.nima.2017.11.009}
  {\path{doi:https://doi.org/10.1016/j.nima.2017.11.009}}.

\bibitem{Fomos:IndexOfRefractionGAGG}
N.~S. Kozlova, et~al., {Optical characteristics of single crystal
  Gd$_3$Al$_2$Ga$_3$O$_{12}$ : Ce}, Modern Electronic Materials 4~(1) (2018)
  7--12.
\newblock \href {https://doi.org/10.3897/j.moem.4.1.33240}
  {\path{doi:10.3897/j.moem.4.1.33240}}.

\end{thebibliography}

\end{document}